\documentclass{article} 
\usepackage{amsmath,amsfonts,amsthm,amssymb}
\usepackage{fullpage}
\usepackage{graphicx}
\usepackage{wrapfig}

\DeclareMathOperator{\Exp}{Exp}

\newtheorem{lemma}{Lemma}

\newtheorem{definition}{Definition}
\newtheorem{corollary}{Corollary}

\newcommand{\C}{{\Bbb C}}

\DeclareMathAlphabet{\varmathbb}{U}{bbold}{m}{n}
\newcommand{\one}{\varmathbb 1}

\newcommand{\ket}[1]{\left| #1 \right\rangle}
\newcommand{\bra}[1]{\left\langle #1 \right|}
\newcommand{\inner}[2]{\left\langle #1 \!\mid\! #2 \right\rangle}

\newcommand{\CG}{\C[G]}

\newcommand{\tr}{\textbf{tr}\,}

\newcommand{\abs}[1]{\left|#1\right|}
\newcommand{\norm}[1]{\left\|#1\right\|}
\newcommand{\U}{\textsf{U}}

\newcommand{\remove}[1]{}

\newcommand{\Z}{{\mathbb Z}}

\newcommand{\e}{{\rm e}}

\newcommand{\Reg}{\textrm{Reg}}

\newcommand{\vrho}{{\boldsymbol{\rho}}}
\newcommand{\vsigma}{{\boldsymbol{\sigma}}}

\newcommand{\vc}{{\boldsymbol{c}}}
\newcommand{\vh}{{\boldsymbol{h}}}

\newcommand{\wg}{\widehat{G}}

\newcommand{\Planch}{{\rm Planch}}
\newcommand{\Span}{{\rm span}}

\newcommand{\ve}{\left( \! \begin{array}{c}}
\newcommand{\ctor}{\end{array} \! \right)}
\newcommand{\mat}{\left( \! \begin{array}{rr}}
\newcommand{\rix}{\end{array} \! \right)}

\newcommand{\rank}{\textbf{rk}\;}

\title{Explicit Multiregister Measurements for Hidden Subgroup Problems} 

\author{Cristopher Moore\\ 
  University of New Mexico\\
\and 
Alexander Russell\\ 
University of Connecticut \\
}

\date{}

\begin{document}
\maketitle 

\begin{abstract}  
  We present an explicit measurement in the Fourier basis that solves
  an important case of the Hidden Subgroup Problem, including the case
  to which Graph Isomorphism reduces.  This entangled measurement uses
  $k=\log_2 |G|$ registers, and each of the $2^k$ subsets of the
  registers contributes some information.  While this does not, in general, 
  yield an efficient algorithm, it generalizes the relationship between Subset Sum 
  and the HSP in the dihedral group, and sheds some light on how quantum algorithms 
  for Graph Isomorphism might work.
\end{abstract}

\section{Introduction: The Hidden Subgroup Problem}

Many problems of interest in quantum computing can be expressed as, or reduced to, an instance of the \emph{Hidden Subgroup Problem} (HSP).  We are given a group $G$ and a function $f$ with the promise that, for some subgroup $H \subseteq G$, $f$ is invariant precisely under translation by $H$: that is, $f(g_1) = f(g_2)$ if any only if $g_1 = g_2 h$ for some $h \in H$.  We then wish to determine the subgroup $H$.  Every known efficient algorithm for this problem---and, indeed, almost every quantum algorithm that provides an exponential speedup over the best known classical algorithm---uses the approach of \emph{Fourier sampling} \cite{BernsteinV93}.  
By preparing a uniform superposition over the elements of $G$, querying the function $f$, and then measuring the value of $f$, we obtain a uniform superposition over one of the (left) cosets of $H$, 
$$
\ket{cH}
= \frac{1}{\sqrt{|H|}} \; \sum_{h \in H} \ket{ch} 
$$
where $c$ is a uniformly random element of $G$.   Alternately, we can view this as a mixed state over the left cosets, the \emph{coset state}, with density matrix
\[ \rho = \frac{1}{|G|} \sum_{c \in G} \ket{cH} \bra{cH} \enspace . \]
We then carry out the quantum Fourier transform on $\ket{cH}$, or equivalently $\rho$, and measure the result.

For example, in Simon's problem~\cite{Simon97}, $G=\Z_2^n$ and there is some $y$ such that $f(x) = f(x+y)$ for all $x$; in this case $H=\{0,y\}$ and we wish to identify $y$.  In Shor's factoring algorithm~\cite{Shor97} $G$ is the group $\Z_n^*$ where $n$ is the number we wish to factor, $f(x) = r^x \bmod n$ for a random $r < n$, and $H$ is the subgroup of $\Z_n^*$ whose index is the multiplicative order of $r$.  
(However, since $|\Z_n^*|$ is unknown, we actually perform the Fourier transform over $\Z_q$ for some
$q=O(n^2)$; see \cite{Shor97} or \cite{HalesH99,HalesH00}.)  
% CM: added this point back in, since the referee on our SICOMP paper objected.
In both these algorithms, $G$ is abelian, and it is not hard to see that for any abelian group a polynomial number\footnote{Throughout the paper, the terms polynomial, subexponential, etc. refer to a function of $\log |G|$.} of experiments of this type allow us to determine $H$.  In essence, each experiment yields a random element of the dual space $H^\perp$ perpendicular to $H$'s characteristic function, and as soon as these elements span $H^\perp$ we can determine a set of generators for $H$ by linear algebra.

While the \emph{nonabelian} hidden subgroup problem appears to be much more difficult, solving it would provide enormous benefits.  In particular, solving the HSP for the symmetric group $S_n$ would provide an efficient quantum algorithm for the Graph Automorphism and Graph Isomorphism problems (see e.g.~\cite{Jozsa00} for a review).  Let $G_1, G_2$ be two rigid, connected graphs of size $n$, and let $H \subset S_{2n}$ be the automorphism group of their disjoint union.  If $G_1 \cong G_2$, then $H=\{1,m\}$ is of order 2, consisting of the identity and an involution $m$ composed of $n$ disjoint transpositions; if $G_1 \not\cong G_2$, then $H$ is the trivial subgroup consisting only of the identity.  Thus even distinguishing subgroups of order 2 from the trivial subgroup would be sufficient to solve this case of Graph Isomorphism.  Other important motivations include the relationship 
between the HSP on the dihedral group and hidden shift problems~\cite{vanDamHI03} and cryptographically important cases of the Shortest Lattice Vector problem~\cite{Regev}.

So far, explicit polynomial-time quantum algorithms for the HSP are known only for a few families of nonabelian groups~\cite{BaconCvDHeisenberg,FriedlIMSS02,GrigniSVV01,HallgrenRT00,InuiLeGall,IvanyosMS01,MooreRRS04,RoettelerB98}.  
However, the basic idea of Fourier sampling can certainly be extended
to the nonabelian case.  Fourier basis functions are homomorphisms
$\phi:G \to \C$ such as the familiar $\phi_k(x) = \e^{2\pi i kx/n}$ when $G$
is the cyclic group $\Z_n$.  In the nonabelian case, one instead
considers \emph{representations} of $G$, namely homomorphisms $\sigma:G \to \U(V)$ where $\U(V)$ is the group of unitary matrices acting on some vector space $V$ of dimension $d_\sigma$.  The \emph{irreducible} representations are those which are not isomorphic to direct sums of representations on lower-dimensional subspaces, and we denote the set of irreducibles by $\wg$.  
%Up to isomorphism, it turns out that the number of irreducible representations equals the number of conjugacy classes of $G$.  Moreover, once a basis for each $V$ is chosen, the matrix elements $\sigma_{ij}$ provide an orthogonal basis for $\CG$, and, in particular, $\sum_\sigma d_\sigma^2 = |G|$.  
We refer the reader to~\cite{FultonH91} for an introduction.  
We denote the set of functions $\psi:G \to \C$ with $\norm{\psi}^2=1$, i.e., 
the Hilbert space of a group-valued register, as $\CG$; then
the quantum Fourier transform consists of transforming vectors in
$\CG$ from the basis $\{ \ket{g} \mid g \in G \}$ to the basis
$\ket{\sigma,i,j}$ where $\sigma$ is the isomorphism type, or ``name,'' of an
irreducible representation and $1 \leq i, j \leq d_\sigma$ index a row and
column (in a chosen basis for $V$).  This transformation can 
be carried out efficiently for a wide variety of groups~\cite{Beals97,Hoyer97,MooreRR04}.
%, although since $d_\sigma$ is typically exponentially large, the choice of basis for $V$ can be crucial to give the matrices $\sigma(g)$ a sparse, highly structured form.  
%The question is then how much information measuring the state $\rho$ in
%this Fourier basis gives us.
%% ACR Minor wording changes above. ``we'' --> ``one''. We don't
%% define $\CG$---oh well.

Several varieties of measurement in the
Fourier basis have been proposed.  \emph{Weak Fourier sampling}
consists of measuring just the name $\sigma$ of the irreducible
representation.  \emph{Strong Fourier sampling} consists of measuring
the name $\sigma$ and the column $j$ in a basis of our choice.  (As the state $\rho$ is mixed
uniformly over the left cosets, it is easy to show that measuring the
row provides no information).  As an
intermediate notion, one can also consider measuring the column in a
\emph{random} basis for $V$.

Unfortunately, a series of negative results have shown that these
types of measurement will not succeed in solving the Hidden Subgroup
Problem in the cases we care most about---in particular, the case
relevant to Graph Isomorphism~\cite{HallgrenRT00,GrigniSVV01,KempeS}.
In particular, Moore, Russell and Schulman~\cite{MooreRS} showed that strong Fourier sampling fails, in the sense that we need an exponential number of experiments on single coset states to distinguish the order-2 subgroups from the trivial subgroup.

However, there is still reason for hope.  In the above description of
Fourier sampling, $f$ is queried just once, giving a coset state on a
single group-valued register.  One can also consider {\em
  multiregister} experiments, in which we carry out $k$ queries of
$f$, prepare $k$ independent coset states, and then perform a joint
measurement on the product state $\vrho = \rho^{\otimes k} = \rho \otimes \cdots \otimes \rho$.
Note that this measurement does not generally consist of $k$
independent measurements; rather, it is an \emph{entangled}
measurement, in which we measure vectors in $\C[G^k]$ along a basis
whose basis vectors are not tensor products of $k$ basis vectors in
$\C[G]$.  For instance, Ip~\cite{Ip} showed that the optimal
  measurement in the dihedral group is already entangled in the
  two-register case.

In one sense we already know that such a measurement can succeed.  Ettinger, H{\o}yer and Knill~\cite{EttingerHK99} showed that the density matrices $\vrho$ become nearly orthogonal for distinct subgroups for some $k = O(\log |G|)$.  As a consequence, a measurement exists which determines the hidden subgroup with high probability.  In~\cite{EttingerHK04} they make this result somewhat more constructive by giving an algorithm which solves the HSP by performing a brute-force search through the subgroup lattice of $G$; however, for groups of interest such as the symmetric groups, this algorithm takes exponential time.  For the dihedral groups in particular, Kuperberg~\cite{Kuperberg03} devised a subexponential algorithm, which uses $2^{O(\sqrt{\log n})}$ time and registers, that works by combining two registers at a time and decomposing them into irreducibles.

Regev~\cite{Regev} provided a beautiful kind of worst-case to average-case quantum reduction, by showing that the HSP for the dihedral group $D_n$ can be reduced to uniformly random instances of the Subset Sum problem on $\Z_n$.  Bacon, Childs, and van Dam~\cite{BaconCvD} deepened this connection by determining the \emph{optimal} multiregister measurement for the dihedral group, and showing that it consists of the so-called \emph{pretty good measurement} (PGM); they used this to show a sharp threshold at $k=\log_2 n$ for the number of registers needed to solve the HSP.
%, and point out that a subexponential-time quantum algorithm can be derived directly from a classical algorithm for moderately dense random cases of Subset Sum due to Flaxman and Przydatek~\cite{FlaxmanP}.  
Moore and Russell~\cite{MooreR::PGM} generalized their results to some extent, showing that the PGM is optimal for arbitrary groups $G$ in the single-register case whenever we wish to distinguish the conjugates of some subgroup $H$ from each other, and optimal in the multiregister case whenever $(G,H)$ form a Gel'fand pair.

Whether a similar approach can be taken to the symmetric group $S_n$ is a major open question.  In particular, we would like to know whether there is a worst-case to average-case reduction analogous to Regev's, connecting the HSP to some Subset Sum-like problem 
%---perhaps involving Young diagrams, which label the irreducible representations of $S_n$---
and whether this would result in new subexponential-time quantum algorithms for Graph Isomorphism.  Some recent results show that indeed any such measurement requires a high degree of entanglement:  Moore and Russell~\cite{MooreR} showed that performing strong Fourier sampling on two registers in $S_n$ requires a superpolynomial number of experiments (specifically, $e^{\Omega(\sqrt{n}/\log n)}$) to distinguish order-2 subgroups $\{1,m\}$ from the identity, or from each other, and conjectured that $\Omega(n \log n)$ registers are necessary.  Hallgren, Moore, R\"otteler, Russell and Sen~\cite{Hallgrenetal} proved this conjecture, showing that $\Theta(n \log n)$ registers are necessary and sufficient.  Interestingly, the variance over $m$ in the observed probability distribution in the multiregister case has a term for each subset of the registers~\cite{MooreRS}, pointing towards an algorithm that finds a subset with particularly high variance, and thus gives a large amount of information about the hidden subgroup.

\paragraph{Our contribution.}  In this paper, we consider the special case of the HSP relevant to Graph Isomorphism: namely, where we wish to distinguish the conjugates of some subgroup $H$ from the trivial subgroup, where $H$ has a ``missing harmonic'' (defined below).  We give an explicit $k$-register measurement in the Fourier basis that distinguishes these two cases.  Our approach relies on decomposing the tensor product of the representations observed in a given subset of the registers into a direct sum of irreducibles.  Each subset of the registers contributes a small amount of information, so that when $k \geq \log_2 |G|$ the measurement succeeds with constant probability.  We hope that this may lead to worst-case to average-case quantum reductions involving generalizations of the Subset Sum problem.

\section{Missing Harmonics}
\label{sec:missing}

We start by preparing independent coset states in $k$ independent $G$-valued registers, giving the tensor product
\begin{equation}
\label{eq:rhok}
 \vrho = \rho^{\otimes k} = \left( \frac{1}{|G|} \sum_{c \in G} \ket{cH} \bra{cH}  \right)^{\!\otimes k} 
= \frac{1}{|G|^k} \sum_{\vc \in G^k} \ket{\vc H^k} \bra{\vc H^k} \enspace . 
\end{equation}
Note that $\vrho$ can also be thought of 
%thought of either as a tensor product of $k$ independent states, each of which is a random left coset of $H \subset G$, or 
as a random left coset of the product subgroup $H^k \in G^k$.  Note also that $\vrho$ is the completely mixed state over $\C[G]^{\otimes k} = \C[G^k]$ if $H$ is the trivial subgroup $\{1\}$.

Here we give an explicit measurement in the Fourier basis which solves an important special case of the HSP, including the case relevant to Graph Isomorphism: namely, given a (non-normal) subgroup $H \subset G$, we wish to distinguish the conjugates of $H$ from the trivial subgroup.  
%As stated above, in Graph Isomorphism, $G=S_{2n}$ and $H$ is the order-2 subgroup generated by the product of $n$ disjoint transpositions.  
Our measurement succeeds with constant probability whenever $k \geq \log_2 |G|$.

Recall that for any representation $\tau$, the average of $\tau$ over a
subgroup $H$ is a projection operator, which we denote $\tau(H) = (1/ |H|
) \sum_{h \in H} \tau(h)$.  Note that $\tau(H)$ is generally not of full rank,
and indeed $\tau(H) = \one_{d_\tau}$ if and only if $H$ is contained in the
kernel of $\tau$.  Let us say that an irreducible representation $\eta$ is a
\emph{missing harmonic} of $H$ if $\eta(H) = 0$; this is then true for
all of $H$'s conjugates as well.  For instance, if $G$ is the dihedral
group $D_n$ and $H=\{1,m\}$ where $m$ is one of the $n$ ``flips,'' then
the sign representation $\pi$ is a missing harmonic.  Similarly, if $G$
is the symmetric group $S_{2n}$ where $n$ is odd and $H$ is the
order-2 subgroup corresponding to an isomorphic pair of rigid graphs,
then the sign representation $\pi$, which takes even and odd
permutations to $+1$ and $-1$ respectively, is a missing harmonic.

For simplicity, we focus on the case where $H$ has some missing harmonic $\eta$; the idea is that if we ever observe it, then we know that the hidden subgroup must be trivial rather than a conjugate of $H$.  The following lemma 
%, proved in Appendix~\ref{app:missing}, 
gives some sufficient conditions for $H$ to have a missing harmonic; these are intended as examples, and are by no means exhaustive.
\newpage

\begin{lemma}
\label{lem:missing}
If any of the following conditions hold, then $H$ has a missing harmonic: 
\begin{enumerate}
\item $H$ is normal and nontrivial.
\item $H$ intersects every coset of some proper normal subgroup $K \lhd G$.
\item $G=S_n$ and $H$ is transitive.
\item $|G|/|H| < C$ where $C = \sum_{\tau \in \wg} d_\tau$.  
\end{enumerate}
\end{lemma}

\begin{proof}  
1) Recall that if $H$ is normal then for every $\tau \in \wg$, either $\tau(H) = \one_{d_\tau}$ or $\tau(H)=0$.  If $H$ is not the trivial subgroup, then the latter must be true for at least one $\tau$.

2) Recall that any irreducible representation of $G/K$ gives an
irreducible representation $\tau$ of $G$ by composing it with the
homomorphism $\phi: G \to G/K$.  Since $\phi(H) = G/K$, we have $\tau(H) = 0$ for
any such $\tau$ other than the trivial representation.  (For instance, in
Graph Isomorphism where $n$ is odd, $H$ is transverse to the
alternating group $A_n$ and $\tau$ is the sign representation.)

3) Let $\tau$ be the \emph{standard representation}, corresponding to the Young diagram $(n-1, 1)$.  This permutes the $n$ vertices of an $(n-1)$-dimensional simplex centered at the origin.  If $H$ is transitive, then for any $i, j$ exactly $1/n$ of the elements of $H$ take vertex $i$ to vertex $j$, and so the average $\tau(H)=0$ is zero.

4) Recall that the \emph{regular representation} $\Reg$, namely $\CG$ under left multiplication by $G$, consists of $d_\tau$ copies of each $\tau \in \wg$.  It is a simple exercise to show that $\rank \Reg(H)$ is the index $|G|/|H|$, and since 
\[ \rank \Reg(H) = \sum_{\tau \in \wg} d_\tau \rank \tau(H) \]
we have $\rank \Reg(H) \geq C$, a contradiction, if $\tau(H) \neq 0$ for all $\tau \in \wg$.
\end{proof}

\section{Decomposing Subsets of the Registers}
\label{sec:subsets}

The state $\vrho$ is a density matrix defined on the Hilbert space
$\C[G^k] = \C[G]^{\otimes k}$.  Since it is completely mixed over left
cosets of $H^k$, it commutes with left multiplication in $G^k$.  It
follows from Schur's lemma~\cite{Kuperberg03,MooreRS,MooreR} that
$\vrho$ is block-diagonal in the Fourier basis, where each block
corresponds to one of the irreducible representations of $G^k$.  These
are tensor products of irreducible representations of $G$, $\vsigma =
\sigma_1 \otimes \cdots \otimes \sigma_k$.  To put it differently, the optimal measurement
is consistent with first performing weak Fourier sampling on each of
the $k$ registers, observing the representation names $\sigma_1,\ldots,\sigma_k$.

The question is how to refine this measurement further, decomposing
$\vsigma$ into smaller subspaces.  (We abuse notation by identifying
subspaces with the name of the representation that acts on them.)
Happily, there is a natural way to do this that respects the
structure of $G$: specifically, we treat $\vsigma$ as a representation
of $G$ (rather than of $G^k$) by restricting to the \emph{diagonal
  action}, where the element $g \in G$ acts by $\vsigma(g) = \sigma_1(g) \otimes
\cdots \otimes \sigma_k(g)$. We can then further decompose $\vsigma$ into
irreducible representations $\tau \in \wg$ under this action.  If we
observe a missing harmonic $\eta$ under this decomposition, we know
that the hidden subgroup is trivial rather than being a conjugate of $H$.  
Unfortunately, in all cases of interest $\eta$ is very low-dimensional (indeed, one-dimensional), 
and so  the chances of observing $\eta$ are exponentially small even
if the hidden subgroup is trivial.  Thus this direct approach does not work.
%% ACR Minor language and typesetting changes above.

Instead, we focus on some subset $I \subseteq [k]$ of the registers.  First, we can decompose $\vsigma$ into the tensor product of the registers inside and outside $I$, $\vsigma
= \left( \bigotimes_{i \in I} \sigma_i \right) \otimes \left( \bigotimes_{i \notin I} \sigma_i \right)$.
Now consider the decomposition of the registers in $I$ into
irreducible representations of $G$ under the diagonal action, in which
we right-multiply\footnote{We use right multiplication because left cosets of $H$ are invariant under right multiplication by $H$.} every register in $I$ by $g$ and leave the other
registers fixed.  We write
%% ACR Minor change just above + ``we write''
$\bigotimes_{i \in I} \sigma_i = \tau_1 \oplus \cdots \oplus \tau_\ell$.
Fixing our missing harmonic $\eta$, for each nonempty $I$ this gives us a subspace 
\[ W^I_{\eta,\vsigma} = 
\left( \bigoplus_{i : \tau_i \cong \eta} \tau_i \right) \otimes \left( \bigotimes_{i \notin I} \sigma_i \right)
\enspace , \] 
and we define $\Pi^I_{\eta,\vsigma}$ as the projection operator which projects onto this subspace.  That is, $\Pi^I_{\eta,\vsigma}$ projects the registers in $I$ into irreducible subspaces isomorphic to $\eta$, and leaves the other registers fixed.  The following lemma 
%, proved in Appendix~\ref{app:pieta}, 
shows that if $\eta$ is a missing harmonic for $H$, then each of these projection operators annihilates  $\vrho$.  

\begin{lemma}  
\label{lem:pieta} 
Suppose that $\eta(H) = 0$.  Then for all nonempty $I \subseteq [k]$ and all $\vsigma$, $\Pi^I_{\eta,\vsigma} \,\vrho = 0$.
\end{lemma}

\begin{proof}  The state $\vrho$ is symmetric under right multiplication by any $\vh \in H^k$.  In particular, it is symmetric under right diagonal multiplication by any $h \in H$ on the registers in $I$.  Let $R^I_H$ be the operator which symmetrizes over this action: that is, the average over all $h \in H$ of the unitary operator that right-multiplies by $\vh$ where $h_i=h$ for $i \in I$ and $h_i=1$ for $i \notin I$.  Then 
\[ \vrho = R^I_H \vrho \,(R^I_H)^\dagger
\quad\mbox{ and }\quad 
\Pi^I_{\eta,\vsigma} R^I_H = 0 \]
and so 
\[ \Pi^I_{\eta,\vsigma} \,\vrho = (\Pi^I_{\eta,\vsigma} R^I_H) \vrho (R^I_H)^\dagger = 0 \enspace . \]
\end{proof}

Now we patch these operators together to form our measurement.  Let 
\[ W_{\eta,\vsigma} = \Span_{I \subseteq [k]} W^I_{\eta,\vsigma} \]
be the span of all these subspaces, and let $\Pi_{\eta,\vsigma}$ be the projection operator onto $W_{\eta,\vsigma}$.  By Lemma~\ref{lem:pieta}, we know that $\Pi_{\eta,\vsigma} \,\vrho = 0$ whenever $\eta$ is a missing harmonic for the hidden subgroup.  Thus we can distinguish the conjugates of $H$ from the trivial subgroup with a measurement operator that reports ``trivial'' if it observes the subspace $W_{\eta,\vsigma}$, and ``don't know'' if it observes the perpendicular subspace $W_{\eta,\vsigma}^\perp$.  Since $\vrho$ is completely mixed if the hidden subgroup is trivial, the probability that our operator reports ``trivial'' in that case is $\dim W_{\eta,\vsigma} / d_\vsigma$.  We wish to show that if $k \geq \log_2 |G|$, the expectation over $\vsigma$ of this fraction is at least $1/2$, so that our measurement distinguishes the trivial subgroup from conjugates of $H$ with constant probability.  

To calculate this expectation, it is convenient to work in the entire Hilbert space $\C[G^k]$ of the $k$ registers, rather than conditioning on having observed the representation names $\vsigma$.  Recall that the action of $G$ on $\CG$ under (right) group multiplication yields the regular representation $\Reg$, and that $\Reg$ contains $d_\sigma$ copies of each $\sigma \in \wg$.  It follows that the fraction of $\CG$, dimensionwise, consisting of copies of $\sigma$ is $d_\sigma^2 / |G|$.  This fraction is also the probability that we observe the representation name $\sigma$ in a given register when we perform weak Fourier sampling on the completely mixed state, and is called the \emph{Plancherel distribution} $\Planch(\sigma)$.  Similarly, $\C[G^k]$ can be thought of as the regular representation of $G^k$, in which case it contains $d_\vsigma = \prod_i d_{\sigma_i}$ copies of each $\vsigma$, giving the Plancherel distribution $\Planch(\vsigma) = \prod_i \Planch(\sigma_i)$.  Thus we have
\[ \Exp_{\vsigma} \frac{\dim W_{\eta,\vsigma}}{d_{\vsigma}} 
= \sum_{\vsigma} \Planch(\vsigma) \frac{\dim W_{\eta,\vsigma}}{d_{\vsigma}} 
= \frac{\sum_{\vsigma} d_{\vsigma} W_{\eta,\vsigma}}{|G|^k}
= \frac{\dim W_\eta}{|G|^k} \enspace . 
\]
In other words, the expected dimensionwise fraction of $W_{\eta,\vsigma}$ in $\vsigma$ is the total dimensionwise fraction of $W_\eta$ in all of $\C[G^k]$, where
\[ W_\eta = \Span_\vsigma W_{\eta,\vsigma} \enspace . \]
We can also write
\[ W_\eta = \Span_{I \subseteq [k]} W_\eta^I \]
where
\[ W_\eta^I = \Span_\vsigma W_{\eta,\vsigma}^I \]
is the subspace of $\C[G^k]$ spanned by vectors for which, if we decompose the registers in $I$ into $G$-irreducibles, we observe the representation name $\eta$, regardless of what $G^k$-representation $\vsigma$ they lie in. 

We wish to lower bound the fraction of $\C[G^k]$ consisting of $W_\eta$.
First, we ask how much of $\C[G^k]$ consists of each $W_\eta^I$.  Recall
that for any representation $\phi$, $\phi \otimes \Reg$ consists of the direct sum
of $d_\phi$ copies of $\Reg$.  In particular, $\Reg^{\otimes \ell}$ contains
$|G|^{\ell-1} d_\sigma$ copies of each $\sigma \in \wg$.  Thus for any $I$, if
$|I|=\ell$ we have
\[ W_\eta^I \cong |G|^{\ell-1} d_\eta \eta \otimes \Reg^{\otimes (k-\ell)} \]
and so
\begin{equation}
\label{eq:trpiw}
 \frac{\dim W_\eta^I}{|G|^k} 
= \frac{|G|^{k-1} d_\eta^2}{|G|^k} 
= \frac{d_\eta^2}{|G|} \enspace . 
\end{equation}
In other words, each $W_\eta^I$ occupies the same fraction of $\C[G^k]$
as $\eta$ occupies of $\C[G]$, namely the Plancherel distribution.  This
is just $1/|G|$ in the cases we care about, but then again there are
$2^k$ subsets $I$.  Our hope is that when $k \sim \log_2 |G|$, then, the
span of all the $W_\eta^I$ occupies a large fraction of the Hilbert
space.

Indeed, if the subspaces $W_\eta^I$ for different $I$ were orthogonal,
their dimensions would simply add, giving $\dim W_\eta = (2^k - 1) \dim
W_\eta^I$; however, it is easy to see (even for $G=\Z_2$ and $k=2$) that
this is not the case.  Instead, it turns out that the subspaces
$W_\eta^I$ have a remarkable statistical property akin to pairwise
independence, which we describe in the next section.

\section{Independent Subspaces}

We say that two subspaces $W_1, W_2$ are \emph{independent} if the
expected squared projection of a random vector $v \in W_1$ into $W_2$ is
just what it would be if $v$ were a random vector in the entire space,
i.e., the dimensionwise fraction of that space occupied by $W_2$.
This is a kind of statistical independence between the events that we
observe $W_1$ and $W_2$ (although if their projection operators do not 
commute, we cannot consider these simultaneously as quantum observables!) 
Formally:
\begin{definition} Let $V$ be a vector space.  Let $W_1, W_2$ be
  subspaces with projection operators $\Pi_1, \Pi_2$.  Let $w$ be chosen
  uniformly at random from the vectors in $W_1$ with norm $1$.  Then
  $W_1$ and $W_2$ are \emph{independent} if
  \[ \Exp_{w} \abs{\Pi_2 w}^2 = \frac{\dim W_2}{\dim V} \enspace . \]
  Equivalently, if $v$ is chosen uniformly at random from the vectors
  in $V$ with norm $1$,
  \[ \Exp_v \abs{\Pi_1 \Pi_2 v}^2 = \frac{\dim W_1}{\dim V} \frac{\dim
    W_2}{\dim V} \enspace . \] A family of subspaces $W_1, \ldots, W_m$ is
  independent if $W_i$ and $W_j$ are independent for any distinct
  $i,j$.
\end{definition}
Note that this definition remains the same if we choose $w$ or $v$
uniformly from an orthonormal basis for $W_1$ or $V$ respectively,
rather than from the sphere of radius $1$.  Indeed, since $\Exp_v
\abs{\Pi_1 \Pi_2 v}^2 = \tr \Pi_1 \Pi_2 / \dim V$, a more compact definition
is the following:
\begin{equation}
\label{eq:trp1p2}
\tr \Pi_1 \Pi_2 = \frac{\tr \Pi_1 \,\tr \Pi_2}{\dim V} = \frac{\dim W_1 \dim W_2}{\dim V} \enspace . 
\end{equation}
Note, however, that $\Pi_1 \Pi_2$ is not a projection operator unless $\Pi_1$ and $\Pi_2$ 
commute.

\begin{lemma} 
\label{lem:ind}
Let $I, J \subseteq [k]$ be distinct and nonempty.  Then $W_\eta^I$ and $W_\eta^J$ are independent.
\end{lemma}

\begin{proof}
  We use the fact that for any representation $U$ and an irreducible
  representation $\tau$, the projection operator onto the \emph{isotypic
    subspace} corresponding to $\tau$---that is, the span of all the
  copies of $\tau$ in $U$---is
  \[
  \Pi_\tau = \frac{d_\tau}{|G|} \sum_{g \in G} \chi_\tau(g)^* \,U(g) \enspace . 
  \] 
  In particular, if $\Pi_\eta^I$ projects onto $W_\eta^I$, we have
  \[ 
  \Pi_\eta^I 
  = \frac{d_\eta}{|G|} \sum_{g \in G} \chi_\eta(g)^* 
  \,\Reg(g)^{\otimes I} \otimes \one^{\otimes([k] \setminus I)} 
  \]
  and 
  \[ \Pi_\eta^I \Pi_\eta^J
= \left( \frac{d_\eta}{|G|} \right)^{\!2} 
\sum_{g,g' \in G} \chi_\eta(g)^* \,\chi_\eta(g')^* 
\,\Reg(g)^{\otimes (I \setminus J)} 
\otimes \Reg(g g')^{\otimes (I \cap J)} 
\otimes \Reg(g')^{\otimes (J \setminus I)} 
\otimes \one^{\otimes ([k] \setminus (I \cup J))} 
\enspace, 
\]
where the notation $A^{\otimes I}$ denotes the tensor product $\otimes_{i \in I} A$.
Taking traces, since $\chi_\Reg(g) = |G|$ for $g=1$ and $0$ for $g \neq 1$,
whenever $I \neq J$ the summand is zero unless both $g=1$ and $g'=1$.  In
this case the summand is $d_\eta^2 |G|^k$, so
\[ \tr \Pi_\eta^I \Pi_\eta^J
= \left( \frac{d_\eta^2}{|G|} \right)^{\!2} |G|^k\enspace.
\]
Since $\tr \Pi_\eta^I = \tr \Pi_\eta^J = (d_\eta^2 / |G|) |G|^k$ by~\eqref{eq:trpiw}, we see that $\Pi_\eta^I$ and $\Pi_\eta^J$ satisfy~\eqref{eq:trp1p2} and are independent. 
\end{proof}

Finally, we lower bound the dimension of the span of a independent family of subspaces with the following lemma, and show $\dim W_\eta / |G|^k \geq 1/2$ whenever $k \geq \log_2 |G|$.   
%The following lemma is proved in Appendix~\ref{app:span}.

\begin{lemma}  
\label{lem:span}
Let $V$ have dimension $D$, and let $W_1, \ldots, W_m \subset V$ be a independent family of subspaces of dimension $d$, and let $W=\Span_i W_i$.  Then 
\[ \frac{\dim W}{D} \geq 1 - \frac{1}{1+md/(D-d)} \enspace . \]
\end{lemma}

\begin{proof}
Let $\Pi_i$ project onto $W_i$ for each $1 \leq i \leq m$, and consider the operator $M=\sum_{i=1}^m \Pi_m$.  Since $M$ is positive and symmetric, it can be diagonalized, and has nonzero eigenvalues $\lambda_1,\ldots,\lambda_t > 0$ where its rank is $t=\dim W$.  Its trace is 
\begin{equation}
\label{eq:trM}
 \sum_{\ell=1}^t \lambda_\ell = md \enspace . 
\end{equation}
Then using~\eqref{eq:trp1p2}, the Frobenius norm of $M$ is
\begin{align}
\norm{M}^2 = \sum_{\ell=1}^t \lambda_\ell^2 
&= \tr M^\dagger M = \sum_{i,j} \tr \Pi_i \Pi_j 
= md + \sum_{i \neq j} \tr \Pi_i \Pi_j 
%= md + \sum_{i \neq j} \sum_{e_i,e_j} \tr \ket{e_i} \inner{e_i}{e_j} \bra{e_j} \nonumber \\
%&= md + \sum_{i \neq j} \sum_{e_i,e_j} \abs{\inner{e_i}{e_j}}^2 
= md + m(m-1) \frac{d^2}{D} \enspace . 
\label{eq:lamsquared}
\end{align}
On the other hand, by Cauchy-Schwartz we have
\begin{equation}
\label{eq:cs}
 \left( \sum_{\ell=1}^t \lambda_\ell \right)^{\!2} \leq \; t \sum_{\ell=1}^t \lambda_\ell^2 
\end{equation}
Combining~\eqref{eq:trM}, \eqref{eq:lamsquared} and~\eqref{eq:cs} gives 
\[ md \leq t \left(1 + (m-1) \frac{d}{D} \right) \]
and so
\[ \frac{t}{D} 
\geq \frac{md}{D+(m-1)d} 
= 1 - \frac{1}{1+md/(D-d)} 
\enspace . \]
\end{proof}

Applying this to the independent family $\{ W_\eta^I \mid I \subseteq [k], I \neq \emptyset \}$ gives the following corollary.
\begin{corollary} For any $k \geq \log_2 |G|$, we have $\dim W_\eta /
  |G|^k \geq \frac{1}{2}$.
\end{corollary}

\begin{proof}
We have $V=\C[G^k]$, $D=|G|^k$, $m=2^k-1 \geq |G|-1$, and $d/D=d_\eta^2/|G| \geq 1/|G|$.  Thus $md/(D-d) \geq 1$, and Lemma~\ref{lem:span} completes the proof.
\end{proof}

\section{The Representation Kickback Trick}

In this section we show how to efficiently carry out the von
Neumann measurement associated with the subspace $W^I_{\eta}$ 
for a fixed subset $I$ of the registers.  Of course, this does not tell us how to efficiently 
carry out the measurement associated with their span $W_\eta$.

It suffices to consider the space $V = \bigotimes_{i \in I} \sigma_i$, decompose $V =
\bigoplus_{\tau \in \wg} a_\tau \tau$ into irreducible representations of $G$, and
implement the measurement associated with the projection operator
$\Pi_\eta$ that projects onto the space spanned by the $a_\eta$ copies of $\eta$
in this direct sum above.  Our approach is essentially the same
as the ``summand extraction'' of Kuperberg~\cite{Kuperberg03}.

To carry out this measurement, we introduce a new $G$-valued control
register, in which we initially prepare $\ket{G}$, the uniform
superposition over $G$. Treating our state now as an element of $\C[G]
\otimes V$, we apply the \emph{controlled $G$-action} operator:
$\mathcal{M}: \ket{g} \otimes \ket{\phi} \mapsto \ket{g} \otimes \vsigma_I(g^{-1})
\ket{\phi}$, where $\vsigma_I(g^{-1})$ is the unitary operator $\bigotimes_{i \in
  I} \sigma_i(g^{-1})$. Finally, we compute the quantum Fourier transform
on the control register and carry out the measurement (on the control
register only) corresponding to the operators $\Pi_\eta$ and $\one -
\Pi_\eta$, where $\Pi_\eta: \C[G] \to \C[G]$ is the operator that projects
onto the irreducible subspaces isomorphic to $\eta$.

To see why this works, we return our attention to $\C[G] \otimes V$.  
Consider the following two $G$-actions on this space: let $D_h: \ket{g} \otimes
\ket{\phi} \mapsto \ket{hg} \otimes \vsigma_I(h)\ket{\phi}$ apply the group action
to \emph{both} the control register and $V$, and let $L_h: \ket{g} \otimes
\ket{\phi} \mapsto \ket{hg} \otimes \ket{\phi}$ apply the group action only on the
control register. Then observe that $\mathcal{M}\circ D_h = L_h \circ
\mathcal{M}$: evidently, any subspace of $\C[G] \otimes V$ that is
invariant under $D_h$ is (unitarily) transformed by $\mathcal M$ to a
subspace that is invariant under $L_h$. Observe now that if $W \subset V$
is an invariant subspace of $V$ (under $\sigma_I(h)$) that is isomorphic
to $\eta$, then $\ket{G} \otimes W$ is isomorphic to $\eta$ under the action
$D_h$, as $\ket{G}$ is invariant under right multiplication by $G$.
Thus such a space is carried to a $L_h$-invariant space by $\mathcal
M$, still isomorphic to $\eta$.

\remove{
Of course, while this measurement can be applied efficiently for any
fixed subset $I$, it is unclear how to efficiently apply the
measurement corresponding to $W_\eta$, which would, in the case of $S_n$,
solve Graph Isomorphism.
}

\section{Discussion}

We have suggested here a general framework for solving cases of the Hidden Subgroup Problem similar to that relevant to Graph Isomorphism.  Each subset of the registers contributes a small amount of information, and the span of all their associated subspaces distinguishes the trivial subgroup from the conjugates of a non-trivial subgroup with a missing harmonic.  Of course, even though we can project into each of these subspaces efficiently, it is far from clear how to product into their span.  However, we might hope, through some partial measurement, to find an \emph{informative subset}: that is, a subset $I$ for which a large fraction of the state lies in $W_\eta^I$ if the hidden subgroup is trivial.

One approach to finding such a subset is a ``sieve,'' in which we combine states in pairs, project their tensor products into irreducible representations, and try to work our way down towards a missing harmonic.  This builds a tree of tensor products with a missing harmonic at its root, and the chosen subset of registers corresponds to the leaves of the tree.  This idea is not original with us: if we consider the special case of the HSP in the dihedral group $D_n$ where we wish to distinguish the trivial subgroup from the $n$ conjugate subgroups $H=\{1,m\}$ where $m$ is a ``flip,'' this is exactly what Kuperberg's algorithm does.  Namely, it combines two-dimensional representations, decomposing them according to $\sigma_i \otimes \sigma_j \cong \sigma_{i+j} \oplus \sigma_{i-j}$, until we reach $\sigma_0 \cong \one \oplus \pi$.  We then perform a measurement inside $\sigma_0$, and if we ever observe the sign representation $\pi$ we know that the hidden subgroup is trivial. 

Using a somewhat different type of sieve, Alagic, Moore and Russell~\cite{AlagicMR} recently obtained a subexponential-time algorithm for the HSP in groups of the form $G^n$ for finite $G$.  Even though these groups have a simple structure, they are similar to $S_n$ in that most of their irreducible representations are exponentially high-dimensional, and it was shown in~\cite{Hallgrenetal} that solving their HSP requires entangled measurements over $\Omega(n)$ registers.  Whether a similar sieve can work for $S_n$, providing an efficient (even subexponential) quantum algorithm for Graph Isomorphism remains an open question.

\section*{Acknowledgments}  We are grateful to Dorit Aharonov, Andrew Childs, Tracy Conrad, Gabor Ivanyos, Greg Kuperberg, Sally Milius, Rosemary Moore, Martin R\"otteler, Leonard Schulman, Pranab Sen, Douglas Strain and Umesh Vazirani for helpful discussions.  We also thank the organizers of the Banff Workshop on Quantum Computation in September 2004, where some of this work was done.  We gratefully acknowledge the support of the NSF through grants EIA-0218563, CCR-0220070, and CCF-0524613, and ARO contract W911NF-04-R-0009.

\end{document}